\documentclass[%
 reprint,
superscriptaddress,
 amsmath,amssymb,
 aps,
]{revtex4-2}

\usepackage{graphicx}
\usepackage{dcolumn}
\usepackage{array}
\usepackage{multirow} 
\usepackage{placeins}  
\usepackage{bm}
\usepackage{ulem}
\usepackage{float}
\usepackage{physics}
\usepackage{hyperref} 
\usepackage{lipsum}
\usepackage{comment}
\usepackage[dvipsnames]{xcolor} 

\begin{document}

\preprint{APS/123-QED}

\title{Elevator Codes: Concatenation for resource-efficient quantum memory under biased noise}

\author{Peter Shanahan}
\affiliation{Alice \& Bob, 49 Bd du Général Martial Valin, 75015 Paris, France}
\affiliation{School of Informatics, University of Edinburgh, United Kingdom}
\author{Diego Ruiz}%
\affiliation{Alice \& Bob, 49 Bd du Général Martial Valin, 75015 Paris, France}
\affiliation{Laboratoire de Physique de l'École Normale Supérieure, École Normale Supérieure, Centre Automatique et Systèmes, Mines Paris, Université PSL, CNRS, Inria, Paris, France}

\date{\today}

\begin{abstract}
Biased-noise qubits, in which one type of error (\textit{e.g.} $X$- and $Y$-type errors) is significantly suppressed relative to the other (\textit{e.g.} $Z$-type errors), can significantly reduce the overhead of quantum error correction. Codes such as the rectangular surface code or XZZX code substantially reduce the qubit overhead under biased noise, but they still face challenges. The rectangular surface code suffers from a relatively low threshold, while the XZZX code requires twice as many physical qubits to maintain the same code distance as the surface code. In this work, we introduce a 2D local code construction that outperforms these codes for noise biases $\eta \ge 7\times10^{4}$, reducing the qubit overhead by over 50\% at $p_Z=10^{-3}$ and $\eta = 2 \times 10^6$ to achieve a logical error rate of $10^{-12}$. Our construction relies on the concatenation of two classical codes. The inner codes are repetition phase-flip codes while the outer codes are high-rate bit-flip codes enabled by their implementation at the logical level, which circumvents device connectivity constraints. These results indicate that under sufficiently biased noise, it is advantageous to address phase-flip and bit-flip errors at different layers of the coding scheme. The inner code should prioritize a high threshold for phase-flip errors, while the bit-flip outer code should optimize for encoding rate efficiency. In the strong biased-noise regime, high-rate outer codes keep the overhead for correcting residual bit-flip errors comparable to that of the repetition code itself, meaningfully lower than that required by earlier approaches.

\end{abstract}

\maketitle

Quantum error correction (QEC) is widely regarded as essential for achieving fault-tolerant quantum computing and enabling practical quantum algorithms~\cite{beverland2022assessing}. This necessity arises because quantum hardware is particularly sensitive to environmental noise, resulting in computational error rates typically ranging between $1\%$ and $0.1\%$~\cite{acharya2024quantum,moses2023race,bluvstein2024logical}. Thus, to execute a quantum algorithm fault tolerantly, the logical quantum information is encoded non-locally in ensembles of qubits, using a QEC code to protect it from noise sources on physical hardware~\cite{shor1995scheme,steane1996multiple}. Furthermore, these codes must detect and correct errors in a fault-tolerant manner~\cite{shor1996fault}, meaning their implementation must tolerate errors in all operations, including those involved in executing the code itself. A $[[n,k,d]]$ code encodes $k$ logical qubits into $n$ physical qubits and has a distance $d$, meaning it can correct up to $\frac{d-1}{2}$ errors and can detect up to $d-1$ errors. The rate of the code is the number of logical qubits per physical qubit defined as $R=k/n$ and heavily dictates the \textit{qubit overhead}, i.e. the number of physical qubits (including ancillas) per logical qubit of the code.

Quantum computation can be realised on various different hardware platforms including superconducting circuits~\cite{acharya2024quantum}, neutral atoms~\cite{xu2024constant}, trapped ions~\cite{reichardt2024demonstration} or photonics~\cite{alexander2024manufacturable}, each with different noise mechanisms and error rates. Despite this diversity, the digitalized Pauli error model consisting of Pauli-$X$ (bit-flip), Pauli-$Z$ (phase-flip), and Pauli-$Y$ errors captures the essential features of physical noise and provides a useful framework for quantum error correction~\cite{shor1995scheme}. In this work, we consider biased-noise qubits, hardware modalities that have a lower probability of one form of Pauli error than the others~\cite{aliferis2008fault}. While some qubits, such as spin qubits~\cite{steinacker2024300}, exhibit a natural noise bias, others, like cat qubits, can be engineered, notably in superconducting circuits, to possess a strongly biased noise profile~\cite{mirrahimi2014dynamically,lescanne2020exponential,rousseau2025enhancing,reglade2024quantum,grimm2019kerr}. Theoretical proposals have also been made to implement cat qubits in trapped ions~\cite{rojkov2024stabilization} and neutral atoms~\cite{omanakuttan2024fault, kruckenhauser2025dark}, along with some preliminary experimental results~\cite{debry2025error}. One of the most prominent examples is superconducting dissipative cat qubits, which exhibit a very strong noise bias and support the implementation of bias-preserving CNOT gates~\cite{guillaud2019repetition,puri2020bias}, enabling the construction of fault-tolerant quantum processors with significantly reduced overhead~\cite{ruiz2025ldpc,chamberland2022building,gouzien2023performance}.

Cat qubits exhibit a unique biased noise profile, characterized by an exponential suppression of bit-flip error rates ($p_X, p_Y$) at the cost of a phase-flip error rate ($p_Z$) increase that remains only linear. The latest experiments report a macroscopic bit-flip lifetime of 22 seconds alongside a phase-flip time of $1.3 \times 10^{-6}$ seconds~\cite{rousseau2025enhancing}, implying that error correction protocols must be strongly biased toward correcting the far more frequent phase-flip errors. For this reason, we define two different distances of a QEC code, $d_Z$ and $d_X$, corresponding respectively to the distance against Pauli-$Z$ and Pauli-$X$ errors. Qubits with a much lower bit-flip error rate than phase-flip error rate $p_Z \gg p_X, p_Y$ require highly biased QEC codes with $d_Z \gg d_X$ to achieve efficient encodings and desirable logical error rates. Previously, the primary proposals for asymmetric codes included the thin rectangular surface code~\cite{chamberland2022building}, the thin XZZX code~\cite{bonilla2021xzzx}, and the XY codes~\cite{tuckett2018ultrahigh}. While these codes lower the qubit overhead relative to the standard surface code, significant challenges remain. The rectangular surface code still incurs significant overhead in the near-threshold regime of phase-flip error, which is relevant for intermediate-term biased-noise qubit architectures (see, for example, Fig. 8 of Ref.~\cite{chamberland2022building}). The XZZX code benefits from a high threshold when used in its unrotated form, but requires twice as many qubits as the surface code to achieve the same code distance~\cite{bonilla2021xzzx,hann2025hybrid}. The XY code suffers from weak boundary effects and, in practice, performs worse than the last two options~\cite{higgott2023improved}. Recently, bias-tailored fractal qLDPC codes have been proposed, offering promising rate and distance properties~\cite{leroux2025romanesco}. While these results are encouraging, further work is needed to assess their logical performance and to address the challenges of implementing such codes in a two-dimensional layout.

\begin{figure}[t]
    \centering
    \includegraphics[width=\columnwidth]{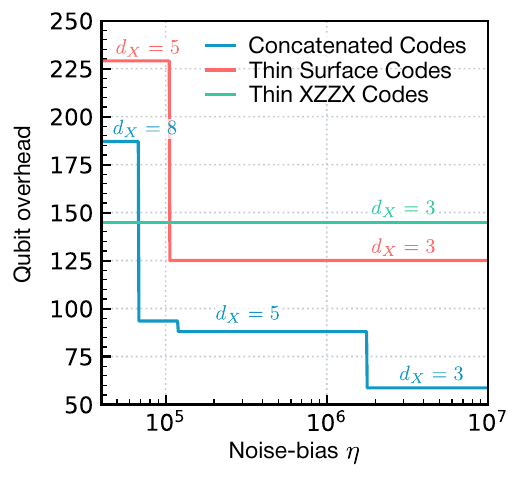}
    \caption{The qubit overhead per logical qubit required at varied noise bias rates to reach a logical error rate of $10^{-12}$ for thin surface codes~\cite{chamberland2022building}, thin XZZX codes~\cite{bonilla2021xzzx} and concatenated repetition codes (this work). The phase-flip error rate is fixed at $p_Z=10^{-3}$ while the noise-bias $\eta$ is varied. The smallest possible code of each type was selected for each noise bias level. For concatenated codes, small changes in overhead are caused by the addition of an extra ancilla inner code block to $\mathcal{C}^{\text{outer}}$. All logical error rates are reported per round of syndrome extraction, using the inner repetition code round for concatenated codes.}
    \label{fig: noise bias}
\end{figure}

We propose the use of concatenated repetition codes for the encoding of qubits with moderate to large noise biases $\eta=\frac{p_Z}{p_X+p_Y}$. These codes are obtained by executing logical operations between blocks of phase-flip repetition code to create a low-distance high-rate bit-flip outer code at the logical level while maintaining only nearest-neighbour physical connectivity. We find that concatenated codes incur lower overheads than thin surface codes and thin XZZX codes for $\eta \geq 7 \times 10^{4}$ and a phase-flip error rate of $p_Z = 10^{-3}$, thanks to the high rate of the bitflip code used  while maintaining the high phase-flip threshold of the repetition code. For instance, at $\eta = 2 \times 10^6$ and $p_Z = 10^{-3}$, achieving a logical error rate of $10^{-12}$ requires a qubit overhead that is about three times lower for concatenated codes compared to the XZZX code, and more than twice as low as that of the thin surface code, as shown in Figure \ref{fig: noise bias}. Concatenated codes also offer an advantage in the near-threshold regime for phase-flip errors, achieving a logical error rate of about $10^{-9}$ at $p_Z = 10^{-2}$ and $\eta = 10^6$ with more than a twofold reduction in qubit overhead compared to the XZZX code and more than a fivefold reduction relative to the thin surface code. Overall, these results indicate that, even when biased-noise qubits fail to achieve sufficiently strong bit-flip suppression, the additional overhead needed to correct bit-flip errors might remain modest for attaining logical error rates pertinent for the first useful quantum algorithms~\cite{beverland2022assessing}. 

\textit{Previous works\ --} Concatenated codes are designed in a two-level structure with an outer code $\mathcal{C}^\text{outer}$ executed using the logical operations of an inner code $\mathcal{C}^\text{inner}$~\cite{knill1996concatenated}. As the encoded logical qubits of $\mathcal{C}^\text{inner}$ are used as the input `physical' qubits of $\mathcal{C}^\text{outer}$, the latter has a far lower input `physical' error rate than hardware provides. The overall distances of the code $\mathcal{C}^\text{combined}$ become the product of the inner and outer code distances $d_X=d_X^\text{inner}d_X^\text{outer}$ and $d_Z=d_Z^\text{inner}d_Z^\text{outer}$. Although code concatenation played a foundational role in QEC~\cite{knill1998resilient, aharonov1997fault} and exhibits high thresholds and rates~\cite{knill2005quantum, yoshida2024concatenate}, topological codes such as the surface code~\cite{kitaev2003fault} are often preferred in practice due to their compatibility with locality constraints in many platforms, including superconducting qubits. However, recent works have shown that code concatenation can outperform a purely surface code based approach by concatenating the surface code with an outer code, while preserving two-dimensional locality~\cite{gidney2025yoked,pattison2025hierarchical}. In this work, we build on the idea of concatenated codes and show that they are particularly effective at correcting a small number of bit-flip errors at the outer code level, rather than addressing them within the inner code. Our scheme, which we refer to as \textit{elevator codes}, consists of the concatenation of two classical codes rather than quantum codes and leverages transversal gate operations at the inner code level.

\begin{figure*}[t]
    \centering
    \includegraphics[width=\textwidth]{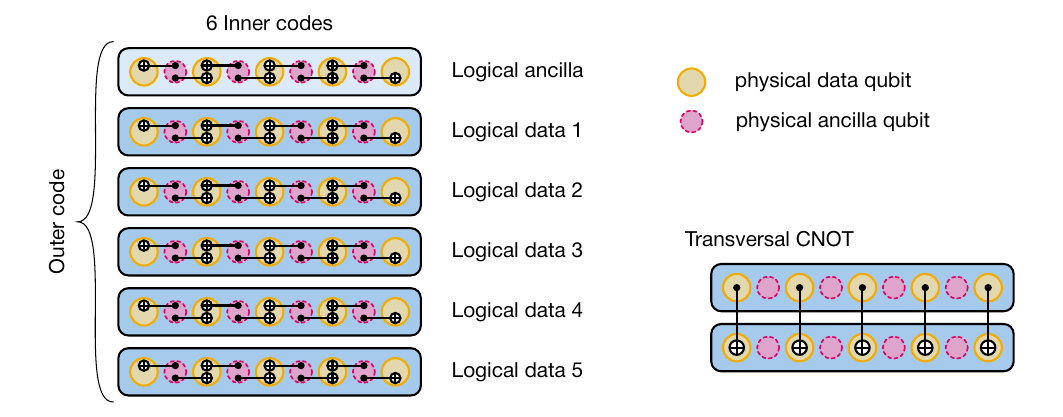}
    \caption{Layout of a concatenated repetition code. $\mathcal{C}^\text{inner}$ are repetition codes and $\mathcal{C}^\text{outer}$ is a $n=5$ qubit code. The 6 blocks of separate $\mathcal{C}^\text{inner}$ codes are the input data and ancilla qubits of one block of $\mathcal{C}^\text{outer}$. Logical CNOT and SWAP gates are executed by transversal CNOT gates to implement the syndrome extraction of $\mathcal{C}^\text{outer}$ using the ancilla logical qubit.}
    \label{fig: schematic}
\end{figure*}

\textit{Overview of code structure\ --} Elevator codes at the inner code level consist of phase-flip repetition code blocks with a variable $d_Z^\text{inner}$ and a fixed $d_X^\text{inner} = 1$. This choice enables a simple connectivity in two dimensions, while benefiting from a high phase-flip threshold, a relatively low-overhead compared to a surface code and the possibility to perform transversal CNOT gates. These transversal CNOT gates are in turn used to implement $\mathcal{C}^\text{outer}$ on a logical level. Additionally, we chose to design the outer code to correct only bit-flip errors, setting $d_Z^\text{outer} = 1$. Indeed, just as stabilizer measurements at the physical level must be carefully designed to prevent error propagation from ancilla to data qubits (commonly known as hook errors), the same caution applies at the logical level. Measuring outer code stabilizers using logical operations from the inner code can similarly introduce logical hook errors. If the outer code has $d_Z^\text{outer} > 1$ and contains high-weight $X$-type checks to correct for inner code logical phase-flip errors, it propagates high-weight logical bitflip hook errors. A solution proposed in Ref.~\cite{gidney2025yoked} involves increasing the bit-flip distance $d_X$ of the ancilla surface code used to measure the check, thereby avoiding the introduction of such hook errors. However, in our case, this would require transforming the logical ancilla repetition codes into thin surface codes, thereby sacrificing the high-threshold and favorable rate properties of the repetition code. Concatenating two classical codes avoids logical hook errors and the need for a thin surface code.

\textit{Implementation of elevator codes\ --} Elevator codes are laid out as a vertical stack of identical repetition code blocks that together form one compact outer block (see Figure~\ref{fig: schematic}). The repetition code blocks are stacked in such a way in order to execute transversal CNOT gates between blocks while adhering to nearest neighbor connectivity. These are the logical operations that will implement $\mathcal{C}^\text{outer}$. This layout enables the measurement of any outer code $Z$-type stabilizer through the following procedure. Each physical data qubit of the ancilla block at the top of the column is initialized in the state $\ket{0}$. A round of repetition code syndrome extraction is then performed on all $\mathcal{C}^\text{inner}$ blocks, projecting the ancilla into the logical state $\ket{0}_L$, up to possible Pauli-$Z$ corrections. The ancilla then undergoes either a CNOT and a SWAP (compiled with two CNOTs) or just a SWAP (compiled with three CNOTs) with the repetition code block directly beneath it (Logical data 1 in Figure \ref{fig: schematic}) depending on the parity check matrix of the outer code. After the first logical operation is completed, another round of inner repetition code syndrome extraction is performed before executing a logical operation between the ancilla and the second data block in the column. The process continues until the first parity check of the outer code is completed. At this point, the ancilla is measured via a transversal $Z$ basis measurement, yielding the value of the syndrome measurement. The ancilla is subsequently reset in place and reinitiates the process to perform the second check of $\mathcal{C}^\text{outer}$. This process is repeated until all checks of the outer code are completed, at which point one round of the outer code is considered executed. This approach, using a single logical ancilla to sequentially measure all checks, minimizes space overhead but can be extended by multiplexing stabilizer measurements with multiple logical ancillae operating in parallel. We note that the presented layout is intended for storing logical qubits and not necessarily for frequent computational access. In practice, a \textit{hot-storage--cold-storage} architecture may be employed, in which a small number of logical qubits (hot storage) are kept in a fully accessible form, while the majority (cold storage) are maintained in a compact layout~\cite{gidney2025yoked,gidney2025factor}. A comprehensive assessment of the impact of such an architecture on quantum computation, in terms of qubit count and time overhead, remains to be carried out.

\textit{Outer code design\ --} While the inner code is restricted to the connectivity of the device, the outer code is executed with logical gates and can therefore relax constraints on code design and open opportunities for high-rate codes with non local stabilizers. We selected three outer codes $\mathcal{C}^\text{outer}$ for our study with parameters $[15,9,3]$, $[15,6,5]$, and $[16,3,8]$. It should be noted that these are not the highest-rate configurations for their given distance and block size, as they were restricted to include only \textit{matchable} codes, in which a single data error triggers at most two stabilizer checks. Matchable codes allow for highly efficient decoding, as the minimum-weight perfect matching algorithm can be applied~\cite{dennis2002topological, Higgott2025sparseblossom}. In our case however, because transversal logical CNOT operations are performed between logical data qubits, the resulting error graph is no longer matchable, and we therefore employed the BP+OSD decoder instead~\cite{roffe2020decoding, Roffe_LDPC_Python_tools_2022, panteleev}. Empirically, we found that the BP+OSD decoder exhibited superior performance on matchable $\mathcal{C}^\text{outer}$ codes. The $[15,9,3]$ code is a simple grid construction similar to~\cite{gidney2025yoked}. The $[15,6,5]$ code was built using the incidence matrix of a Peterson Graph as described in~\cite{zemor2009cayley}. The $[16,3,8]$ code was found empirically by enforcing the maximum degree condition for a matchable graph. While larger codes could increase the rate and decrease the overhead, they were not considered due to simulation constraints. As $\mathcal{C}^\text{inner}$ is a $[d_Z, 1, d_Z]$ phase-flip repetition code and $\mathcal{C}^\text{outer}$ a $[n,k,d_X]$ bit-flip code, the parity check matrix of the resulting $[[nd_Z,k,\{d_X,d_Z\}]]$ quantum code can be expressed as
\begin{equation}
\begin{aligned}
H(\mathcal{C}^\text{combined})=
\begin{bmatrix}
H_X & 0 \\
0 & H_Z
\end{bmatrix}=\\
\begin{bmatrix}
I_n \otimes H(\mathcal C^\text{inner}) & 0 \\
0 & H(\mathcal C^\text{outer})\otimes \mathbf{1}_{1 \times d_Z}
\end{bmatrix}
\end{aligned}
\end{equation}
where $\mathbf{1}_{1 \times d_Z}$ denotes the $1 \times d_Z$ row vector of ones, and $I_n$ is the $n \times n$ identity matrix. The precise structure of all $\mathcal{C}^\text{outer}$ codes can be found in Appendix~\ref{app:parity check matrix}.

\textit{Fault-tolerance and decoding\ --} While executing the outer code, it is essential to ensure that the phase-flip protection provided by the inner code remains intact. In contrast to superconducting surface code architectures, where logical operations typically require lattice surgery and $\mathcal{O}(d)$ rounds, transversal CNOT gates used to execute $\mathcal{C}^\textrm{outer}$ can be performed in $\mathcal{O}(1)$ round, interleaved between each round of repetition code $\mathcal{C}^\textrm{inner}$ stabilizer measurements~\cite{cain2024correlated, zhou2024algorithmic}. Transversal CNOTs, however, transform the decoding graph into a hypergraph, which cannot be straightforwardly decoded using minimum-weight perfect matching. In this work, syndrome information is decoded using a BP+OSD decoder to efficiently handle the hyperedges introduced by transversal CNOTs~\cite{roffe2020decoding}. However, during the development of this work, several studies have proposed adaptations of minimum-weight perfect matching for efficiently decoding syndromes in the presence of transversal CNOTs~\cite{serra2025decoding, cain2025fast, turner2025scalable}. While we do not expect significant changes in the logical error rate compared to BP+OSD, minimum-weight perfect matching offers a substantially faster decoding alternative. We also chose to enforce a minimum of $d_Z$ rounds between the initialization and measurement of the logical ancilla, since the $X$-type stabilizer projection of the repetition code prepared in $\ket{0}^{\otimes d}$ is random and its value must be learned through sufficiently many rounds. With BP+OSD decoding, we observed improved performance under this condition, but the recent decoders may allow this constraint to be relaxed. If the number of logical operations $n_{L}$ (and therefore number of rounds of $\mathcal{C}^\text{inner}$ syndrome extraction) needed to perform a check is such that $n_{L} < d_Z$, $d_Z-n_{L}$ repetition code syndrome extraction rounds are performed while the ancilla is stationary, before it is measured transversally in the $Z$ basis. In practice, for the chosen outer codes and in the operating regime considered, $n_{L} < d_Z$. Consequently, if only a single logical ancilla is used, one round of the outer code corresponds to $m \times d_Z$ rounds of the inner code, where $m$ denotes the number of checks in $H(\mathcal{C}^\text{outer})$. While using one ancilla logical qubit minimizes the number of qubits, it lowers the performance of the outer code as the longer left between code rounds, the more Pauli-$X$ errors will accumulate and the effective physical error rate of $\mathcal{C}^\text{outer}$ will rise. In the simulations performed, we considered either one or two logical ancilla, but in practice, for each target logical error rate, there is an optimal number of logical ancilla qubits to minimize the overhead.

\begin{figure}[htbp]
    \centering
    \includegraphics[width=\columnwidth]{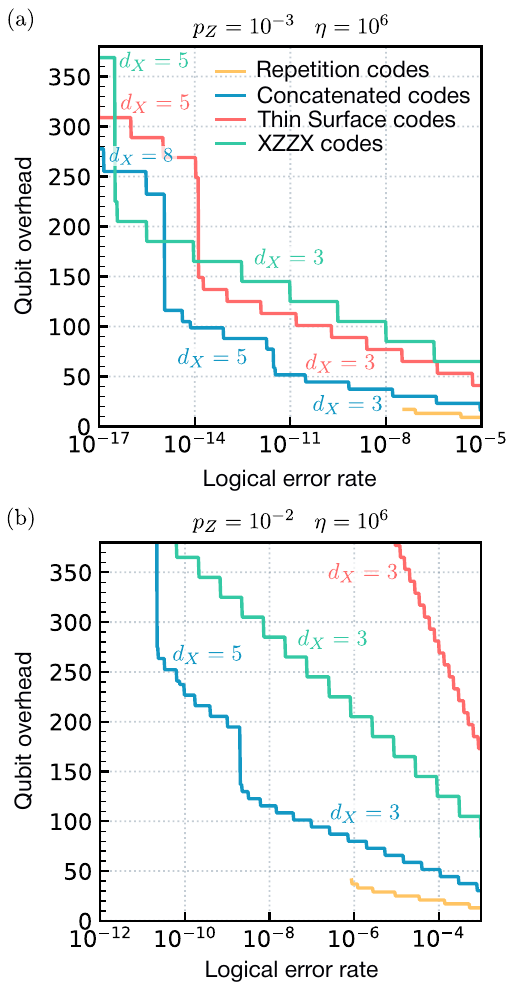}
    \caption{Qubit overhead as a function of the logical error rate with repetition codes~\cite{guillaud2019repetition}, concatenated repetition codes (this work), thin surface codes~\cite{chamberland2022building} and thin XZZX codes~\cite{bonilla2021xzzx}. The qubit overhead includes ancilla qubits and is reported per logical qubit, while the logical error rate is reported per error correction cycle, with the cycle of the inner repetition code used for concatenated codes. (a) Heavily biased error model with deep threshold phase-flip error rates $p_Z$. Concatenated codes offer lower overhead than thin surface codes and thin XZZX codes, and remain effective up to a logical error rate of about $10^{-15}$. While the repetition code has lower overhead, the logical error rates it can achieve are limited by its inability to correct bit-flip errors. (b) Intermediate error regime for medium-term biased-noise qubits. Concatenated codes have a lower overhead than thin surface and thin XZZX codes up to a logical error rate of $10^{-11}$.}
    \label{fig: logical error}
\end{figure}

\textit{Results of numerical simulations\ --} Figure \ref{fig: noise bias} illustrates the qubit overhead (including ancilla qubits) per logical qubit required to achieve a logical error rate of $10^{-12}$ at a physical phase-flip rate of $p_Z = 10^{-3}$, across different values of noise bias $\eta$. The concatenated repetition code architecture demonstrates a favorable qubit overhead than both the thin surface code and the thin XZZX code when $\eta > 7 \times 10^{4}$. In particular, for $\eta \ge 2 \times 10^{6}$, the overhead for concatenated codes is about three times lower than that of the XZZX code and more than twice as low as that of the thin surface code. We note that such high noise biases have already been realized in cat qubit experiments (e.g.,~\cite{rousseau2025enhancing,reglade2024quantum}). Furthermore, theoretical studies of hybrid cat-transmon architectures indicate that achieving a phase-flip rate of $p_Z = 10^{-3}$ with a noise bias on the order of $\eta = 10^{4}$ is feasible with current state-of-the-art coherence times~\cite{hann2025hybrid}, promisingly close to the values where concatenated codes begin to offer a clear overhead advantage.

Figure \ref{fig: logical error}(a) presents the logical error rates achievable at a bias of $\eta=10^6$ and a phase-flip error rate of $p_Z=10^{-3}$ while Figure \ref{fig: logical error}(b) presents a closer term noise model with the same noise bias but a phase-flip error rate of $p_Z=10^{-2}$. In the former case, for logical error rates ranging from $10^{-8}$ to $10^{-15}$, the concatenated codes demonstrate up to a twofold reduction in qubit overhead compared to the thin XZZX and thin surface codes. In the latter case, for logical error rates between $10^{-6}$ and $10^{-11}$, the advantage of concatenated codes becomes even more pronounced, reaching a three-fold improvement. This larger gap arises because the phase-flip threshold of the surface code is roughly five times lower than that of concatenated codes, a difference that becomes increasingly significant as the physical error rate approaches these thresholds. For the XZZX code, since its threshold is roughly equal to that of the concatenated codes, the overhead improvement remains consistent across all physical phase-flip error rates. Interestingly, both plots show that the overhead of concatenated codes is only slightly higher than that of standalone repetition codes, up to logical error rates of $10^{-12}$ for $p_Z = 10^{-3}$ and $10^{-9}$ for $p_Z = 10^{-2}$. This shows that implementing an outer code at the logical level can significantly enhance the achievable logical error rate at little additional cost. This advantage arises from two key factors. First, the outer code possesses a high rate. Second, its execution does not degrade the threshold performance of the inner repetition code. We anticipate that this overhead gap will narrow further with the use of larger, higher rate outer codes $\mathcal{C}^\text{outer}$. Overall, these results reinforce the potential of concatenated coding strategies for biased noise qubits.

\textit{Conclusion and outlook\ --} We have introduced elevator codes, the concatenation of two classical codes with an asymmetric distance profile tailored to biased-noise qubits. We have shown, through extensive simulations, the favourability of concatenated repetition codes over thin surface codes and thin XZZX codes in terms of qubit overhead for noise biases as low as $7 \times 10^{4}$. There are extensive parameters to be tuned further in the optimization of concatenated codes. Investigations into the optimal number of logical ancilla and code layout offer avenues for possible improvements. Moreover, employing larger, higher-rate outer codes could offer further reductions in the overhead of concatenated codes, particularly at lower noise bias levels, and represents a promising direction for future works. They could also further narrow the overhead gap between a repetition code that corrects only phase-flip errors and a concatenated code that also corrects a small number of bit-flip errors. We note that repetition codes of $\mathcal{C}^\text{inner}$ could be replaced by higher rate phase-flip codes~\cite{ruiz2025ldpc}, which reduce qubit overhead while slightly increasing the qubit connectivity and layout. 

We would like to thank Jérémie Guillaud, Mazyar Mirrahimi and Christophe Vuillot for many discussions about the scheme.

\newpage

\bibliographystyle{unsrt}
\bibliography{refs}

\newpage

\appendix

\section{Error model}

The code performance were evaluated using the STIM~\cite{STIM} error correction package. Both $X$- and $Z$-type logical memory simulations were performed and the logical error rate is calculated as $p_{L} = p_{Z_L} + p_{X_L}$. A simple biased-noise model, parameterized by Pauli-$X$ and Pauli-$Z$ type noise with probabilities $p=\{p_X,p_Z\}$, was employed as summarized in Table~\ref{tab:error_model}. $Y$-type errors are not included in the simulations for two reasons. First, in the $Z$-memory simulation, these errors are negligible compared to $Z$ errors within the noise-bias regime considered in this work. Second, in the $X$-memory simulation, accounting for $Y$ errors would require decoding extremely large graphs, which is computationally intractable. We nevertheless expect that including $Y$ errors would only slightly affect the results presented here.

\begin{table}[h]
    \centering
    \newcolumntype{C}[1]{>{\centering\arraybackslash}m{#1}}
    \begin{tabular}{ C{2.5cm}  C{2.5cm}  C{2.5cm} } 
        \hline
        \textbf{Gate} & \textbf{Error} & \textbf{Probability} \\ 
        \hline
        \hline
        \multirow{2}{=}{\centering $\mathcal P_{\ket{0}}, \mathcal M_Z$} & $I$ & $1-p_X$ \\
        & $X$ & $p_X$ \\
        \hline
        \multirow{2}{=}{\centering $\mathcal P_{\ket{+}}, \mathcal M_X$} & $I$ & $1-p_Z$ \\
        & $Z$ & $p_Z$ \\
        \hline
        \multirow{3}{=}{\centering Idle} & $I$ & $1-p_Z-p_X$ \\
        & $Z$ & $p_Z$ \\
        & $X$ & $p_X$ \\
        \hline
        \multirow{7}{=}{\centering CNOT} & $I \otimes I$ & $1-p_X - p_Z$ \\  
        &$I \otimes Z$ & $p_Z/3$ \\  
        &$Z \otimes I$ & $p_Z/3$ \\  
        &$Z \otimes Z$ & $p_Z/3$ \\  
        &$I \otimes X$ & $p_X/3$ \\  
        &$X \otimes I$ & $p_X/3$ \\  
        &$X \otimes X$ & $p_X/3$ \\  
        \hline
    \end{tabular}
    \caption{Circuit-level noise model with error probabilities assigned to each single-qubit gate, two-qubit gate, and idle location in the circuit.}
    \label{tab:error_model}
\end{table}

\section{Simulation modeling and fitting}
\label{app:fit}
 
\textit{Simulation of concatenated codes\ --} Concatenated code circuits are too large to sample their error rate in all parameter regimes presented in this work. In order to derive a model for the logical error rate of concatenated codes, separate $X$-type and $Z$-type memory simulations were performed for various $\mathcal{C}^\text{outer}$ codes and $Z$ distances $d_Z$. The logical error rates in the intractable region were extrapolated from these simulations. In combining both the Pauli-$X$ logical error rate ($p_{X_L}$) and the Pauli-$Z$ logical error rate ($p_{Z_L}$), the overall logical error rate is given by $p_{L} = p_{Z_L} + p_{X_L}$. For the XZZX code and the thin surface code, the logical error rate is reported per round of error correction. For concatenated codes, it is reported per round of the inner repetition code. This provides a fair basis for comparison, since the focus of this work is on quantum memory, where the objective is to preserve logical information for as many rounds as possible.

The $Z$-type memory simulations were performed on the $[15,9,3]$, $[15,6,5]$ and $[16,3,8]$ outer codes combined with inner codes with $Z$ distances of $d_Z={9,11,13,15}$. For the $[15,9,3]$ and $[15,6,5]$ outer codes the simulations were performed for 5 rounds of each outer code, corresponding to $5md_Z$ rounds of the inner code where $m$ is the number of checks in $\mathcal{C}^\text{outer}$, at $p_X$ physical bit-flip error rates in the range $10^{-6}\leq p_X \leq 10^{-5}$. The $[16,3,8]$ code was simulated for 8 outer code rounds corresponding to $8md_Z$ rounds of inner code. Separate parameters for each $\mathcal{C}^\text{outer}$ were fitted to the equation
\begin{equation}
    p_{X_L}(p_X,d_Z) = d_Z^c (a p_X)^b 
    \label{pxl}
\end{equation}
and the results are presented in Table~\ref{tab:outer_code_fit}.

\begin{table}[h]
\centering
\newcolumntype{C}[1]{>{\centering\arraybackslash}p{#1}}
\renewcommand{\arraystretch}{1.2}
\begin{tabular}{ C{8em} | C{4em} | C{4em} | C{4em} } 
  \hline
  \textbf{Outer Code} & \textbf{a} & \textbf{b} & \textbf{c} \\ 
  \hline
  \hline
  [15,9,3] & 37.18 & 1.94 & 2.33 \\
  \hline
  [15,6,5] & 115.14 & 2.76 & 3.73 \\
  \hline
  [15,6,5] & \multirow{2}{*}{88.47} & \multirow{2}{*}{2.86} & \multirow{2}{*}{3.89} \\
  (2 ancillae) & &  &  \\
  \hline
  [16,3,8] &  61.99 & 3.57 & 5.74 \\
  \hline
\end{tabular}
\renewcommand{\arraystretch}{1.0}
\caption{Outer code fit parameters.}
\label{tab:outer_code_fit}
\end{table}

For $X$-type memory experiments, the circuit was simulated for one outer code round corresponding to $md_Z$ inner code rounds. Due to the similar nature of how the outer codes are executed and the time complexity of the simulations, only the $[15,9,3]$ outer code was simulated at inner code distances $d_Z=9,11,13$ and physical error rates in the range $5\times 10^{-3}\leq p_Z \leq 10^{-2}$. The logical error rate per round and per logical qubit was then fitted to
\begin{equation}
    p_{Z_L}(p_Z,d_Z) = 0.12(34.4 p_Z)^{0.94 \frac{d_Z+1}{2}}.
    \label{pxl}
\end{equation}
The power scaling with distance and the threshold are comparable to those of the standalone repetition code, though slightly lower, which we attribute to the suboptimal performance of the BP+OSD decoder.

In order to generalize this model to the $[15,6,5]$ and $[16,3,8]$ codes, two prefactors must be added to account for the number of repetition code blocks used in the outer code $n_b$ and the number of logical qubits encoded in the outer code $k$. Since the $[15,9,3]$ code has $n_b = 16$ and $k = 9$, the general model takes the form
\begin{equation}
    p_{Z_L}(p_Z,d_Z,n_b,k) = \frac{n_b}{16}\frac{9}{k}\times 0.12(34.4p_Z)^{0.94\frac{d_Z+1}{2}}
    \label{pxl}
\end{equation}
This constitutes a reasonable extrapolation, as all outer codes are executed in the same manner.

\textit{Simulation of repetition, surface codes\ --} 
The repetition code was simulated for $X$- and $Z$-type memory experiments using the noise model outlined in Table~\ref{tab:error_model}. The data was collected from experiments consisting of $3d_Z$ rounds for distances $d_Z=3,5,7,9,11$. The resulting error models for are given by
\begin{equation}
    p_{Z_L}(p_Z,d_Z)=0.13(25.02p_z)^{0.99\frac{d_Z +1}{2}}
    \label{pzl_repetition}
\end{equation}
\begin{equation}
    p_{X_L}(p_X,d_Z)=3.88p_Xd_Z
    \label{pzl_repetition}
\end{equation}
Both the rotated thin surface code and the thin XZZX code were simulated using the noise model outlined in Table~\ref{tab:error_model}. $X$- and $Z$-type memory experiments were carried out for 15 rounds for both codes at distances $d_X=3,5$ and $d_Z=5,7,9,11$. 

For $X$-type memory experiments, in order to accurately capture the near threshold behavior of the thin surface code at $p_Z=10^{-2}$, used in Figure \ref{fig: logical error}(b), an additional model was created for this specific error rate. 
The general model used for $X$-type memory experiments was
\begin{equation}
    p_{Z_L}(p_Z,d_Z) = a(b p_Z)^{c\frac{d_Z +1}{2}}.
    \label{pzl_surface}
\end{equation}
Parameters of the model for different codes are found in Table \ref{tab:surface_z_error}.

\begin{table}[h]
\centering
\newcolumntype{C}[1]{>{\centering\arraybackslash}p{#1}}
\renewcommand{\arraystretch}{1.2}
\begin{tabular}{ C{10em} | C{4em} | C{4em} | C{4em} | C{4em}} 
  \hline
  \textbf{Outer Code} & \textbf{$d_X$} & \textbf{a} & \textbf{b} & \textbf{c} \\ 
  \hline
  \hline
  Rotated ($p_Z = 10^{-2}$) & 3 & 0.05 & 4.23 & 0.08 \\
  \hline
  Rotated ($p_Z = 10^{-2}$) & 5 & 0.04 & 2.18 & 0.03 \\
  \hline
  Rotated ($p_Z < 10^{-2}$) & 3 & 0.14 & 72.66 & 0.97 \\
  \hline
  Rotated ($p_Z < 10^{-2}$) & 5 & 0.16 & 84.69 & 0.94 \\
  \hline
  XZZX & 3 & 0.36 & 30.60 & 1.00 \\
  \hline
  XZZX & 5 & 0.64 & 28.92 & 1.01 \\
  \hline

\end{tabular}
\renewcommand{\arraystretch}{1.0}
\caption{$X$-type memory experiment parameters for surface codes.}
\label{tab:surface_z_error}
\end{table}

For $Z$-type memory experiments, the general model used was 
\begin{equation}
    p_{X_L}(p_X,d_Z)=(d_Z)^a(bp_X)^c
    \label{pxl_surface}
\end{equation}
Parameters of the model for different codes are found in Table \ref{tab:surface_x_error}.

\begin{table}[h]
\centering
\newcolumntype{C}[1]{>{\centering\arraybackslash}p{#1}}
\renewcommand{\arraystretch}{1.2}
\begin{tabular}{ C{10em} | C{4em} | C{4em} | C{4em} | C{4em}} 
  \hline
  \textbf{Outer Code} & \textbf{$d_X$} & \textbf{a} & \textbf{b} & \textbf{c} \\ 
  \hline
  \hline
  Rotated  & 3 & 1.06 & 19.30 & 2.00 \\
  \hline
  Rotated  & 5 & 1.19 & 19.30 & 2.97 \\
  \hline
  XZZX & 3 & 1.03 & 0.97 & 1.99 \\
  \hline
  XZZX & 5 & 1.06 & 12.38 & 2.93 \\
  \hline

\end{tabular}
\renewcommand{\arraystretch}{1.0}
\caption{$Z$-type memory experiment parameters for surface codes.}
\label{tab:surface_x_error}
\end{table}

\section{Outer Code Parity Check Matrices}
\label{app:parity check matrix}

This appendix shows the parity check matrices of the outer codes used.

\begin{figure}[H]
\centering
\[
H =
\left[
\begin{array}{ccccccccccccccc}
1 & 0 & 1 & 0 & 1 & 1 & 0 & 0 & 0 & 0 & 0 & 0 & 0 & 0 & 0 \\
0 & 0 & 0 & 0 & 0 & 0 & 1 & 0 & 0 & 1 & 0 & 0 & 1 & 0 & 1 \\
0 & 0 & 0 & 0 & 0 & 1 & 0 & 0 & 1 & 1 & 0 & 0 & 0 & 1 & 0 \\
0 & 0 & 0 & 0 & 1 & 0 & 0 & 0 & 0 & 0 & 1 & 1 & 1 & 0 & 0 \\
0 & 0 & 0 & 1 & 0 & 0 & 0 & 1 & 1 & 0 & 0 & 1 & 0 & 0 & 0 \\
0 & 1 & 1 & 1 & 0 & 0 & 1 & 0 & 0 & 0 & 0 & 0 & 0 & 0 & 0
\end{array}
\right]
\]
\caption{Parity-check matrix of the \([15,9,3]\) code.}
\label{fig:parity-check-15-9-3}
\end{figure}

\begin{figure}[H]
\centering
\[
H =
\left[
\begin{array}{ccccccccccccccc}
1 & 0 & 0 & 0 & 1 & 1 & 0 & 0 & 0 & 0 & 0 & 0 & 0 & 0 & 0 \\
1 & 1 & 0 & 0 & 0 & 0 & 1 & 0 & 0 & 0 & 0 & 0 & 0 & 0 & 0 \\
0 & 1 & 1 & 0 & 0 & 0 & 0 & 1 & 0 & 0 & 0 & 0 & 0 & 0 & 0 \\
0 & 0 & 1 & 1 & 0 & 0 & 0 & 0 & 1 & 0 & 0 & 0 & 0 & 0 & 0 \\
0 & 0 & 0 & 1 & 1 & 0 & 0 & 0 & 0 & 1 & 0 & 0 & 0 & 0 & 0 \\
0 & 0 & 0 & 0 & 0 & 0 & 0 & 0 & 0 & 1 & 1 & 0 & 0 & 0 & 1 \\
0 & 0 & 0 & 0 & 0 & 1 & 0 & 0 & 0 & 0 & 0 & 0 & 1 & 1 & 0 \\
0 & 0 & 0 & 0 & 0 & 0 & 1 & 0 & 0 & 0 & 1 & 1 & 0 & 0 & 0 \\
0 & 0 & 0 & 0 & 0 & 0 & 0 & 0 & 1 & 0 & 0 & 1 & 1 & 0 & 0 \\
0 & 0 & 0 & 0 & 0 & 0 & 0 & 1 & 0 & 0 & 0 & 0 & 0 & 1 & 1
\end{array}
\right]
\]
\caption{Parity-check matrix of the \([15,6,5]\) code.}
\label{fig:parity-check-15-6-5}
\end{figure}

\begin{figure}[H]
\centering
\[
H =
\left[
\begin{array}{cccccccccccccccc}
1 & 0 & 0 & 0 & 1 & 1 & 1 & 0 & 0 & 0 & 0 & 0 & 0 & 0 & 0 & 0 \\
1 & 1 & 0 & 0 & 0 & 0 & 0 & 0 & 0 & 0 & 0 & 0 & 0 & 0 & 0 & 0 \\
0 & 1 & 1 & 0 & 0 & 0 & 0 & 0 & 0 & 0 & 0 & 0 & 0 & 0 & 0 & 0 \\
0 & 0 & 1 & 1 & 0 & 0 & 0 & 0 & 0 & 0 & 0 & 0 & 0 & 0 & 0 & 0 \\
0 & 0 & 0 & 0 & 1 & 0 & 0 & 1 & 0 & 0 & 0 & 0 & 0 & 0 & 0 & 0 \\
0 & 0 & 0 & 0 & 0 & 0 & 0 & 1 & 1 & 0 & 0 & 0 & 0 & 0 & 0 & 0 \\
0 & 0 & 0 & 0 & 0 & 0 & 0 & 0 & 1 & 1 & 0 & 0 & 0 & 0 & 0 & 0 \\
0 & 0 & 0 & 0 & 0 & 1 & 0 & 0 & 0 & 0 & 1 & 0 & 0 & 0 & 0 & 0 \\
0 & 0 & 0 & 0 & 0 & 0 & 1 & 0 & 0 & 0 & 0 & 0 & 0 & 1 & 0 & 0 \\
0 & 0 & 0 & 0 & 0 & 0 & 0 & 0 & 0 & 0 & 1 & 1 & 0 & 0 & 0 & 0 \\
0 & 0 & 0 & 0 & 0 & 0 & 0 & 0 & 0 & 0 & 0 & 1 & 1 & 0 & 0 & 0 \\
0 & 0 & 0 & 0 & 0 & 0 & 0 & 0 & 0 & 0 & 0 & 0 & 0 & 1 & 1 & 0 \\
0 & 0 & 0 & 0 & 0 & 0 & 0 & 0 & 0 & 0 & 0 & 0 & 0 & 0 & 1 & 1
\end{array}
\right]
\]
\caption{Parity-check matrix of the \([16,3,8]\) code.}
\label{fig:parity-check-16-3-8}
\end{figure}

\newpage

\end{document}